\documentclass[a4paper]{elsart}
\usepackage{morefloats}
\usepackage{afterpage}
\usepackage{graphicx}
\usepackage{tabularx}
\usepackage{rotating}
\usepackage{subfigure}
\begin{document}
\begin{frontmatter}
\begin{flushright}
DESY 07-196
\end{flushright}
\title{Study of Micro Pixel Photon Counters for a high granularity scintillator-based hadron calorimeter}
\author[AHHUNI,DESY]{N. D'Ascenzo\thanksref{email}},
\author[DESY]{A. Eggemann},
\author[DESY]{E. Garutti}

\address[AHHUNI]{University of Hamburg, Edmund-Siemers-Allee 1, D-20146, Hamburg, Germany}
\address[DESY]{DESY, Notkestr. 85, D-22607 Hamburg, Germany}
\thanks[email]{Corresponding author: nicola.dascenzo@desy.de}
\begin{abstract}
A new Geiger mode avalanche photodiode, the Micro Pixel Photon Counter (MPPC), was recently released by Hamamatsu. It has a high photo-detection efficiency in the 420~nm spectral region. This product can represent an elegant candidate for the design of a high granularity scintillator based hadron calorimeter for the International Linear Collider. In fact, the direct readout of the blue scintillation photons with a MPPC is a feasible techological solution. The readout of a plastic scintillator by a MPPC, both mediated by the traditional wavelength shifting fiber, and directly coupled, has been systematically studied.
\end{abstract}
\end{frontmatter}

\section*{Introduction}
The prototype of the highly granular hadronic calorimeter  for the International Linear Collider (Calice HCAL \cite{cal2}) consists of finely segmented arrays of scintillators. Each  tile  is read out by a Geiger mode avalanche photodiode\cite{sav1,sipm1,sipm2}  (the SiPM) produced by MEPHI (Moscow Engineering and Physics Institute), whose specific design matches the required performances needed for its application in the calorimeter \cite{cal1}. The small size (1~mm$^{2}$), high gain ($10^{6}$) and relatively low bias voltage (30-70~V) of the SiPM made it the natural candidate for the high granularity calorimeter; the photodetector could be directly installed in the tile. However, the SiPM is a green sensitive photodetector; the coupling with the blue emitting scintillator is obtained via a green wavelength shifting fibre, installed in a groove on the scintillator itself.  A simplification of this elementary structure would be highly desirable, in order to extend the concept to a large scale detector. The new generation of Geiger mode avalanche photodiodes, produced by Hamamatsu, shows a better optical sensitivity in the 420~nm spectral region, making it possible to investigate the direct readout of the scintillation tile. In this study the response to a minimum ionizing particle (mip) of a plastic organic scintillator read out by the most recent MPPC\cite{Hama}, with both 400 and 1600 pixels, was analyzed. The mip signal is in fact a benchmark for the high granularity hadron calorimeter response: it has to be well separated from thermally induced noise. The direct readout option was compared with the established green wavelength shifting mediated design.

\section{The experimental setup}
\begin{figure}
\centering
\includegraphics[scale=0.5]{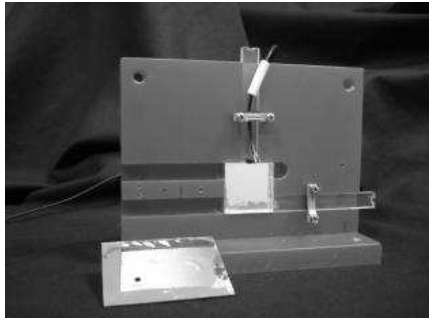}
\label{DR}
\caption{Setup for the direct coupling. The scintillator is installed in a specific housing. The MPPC is fixed at one end of a plastic holder which can be located in two different positions: the coupling is reproducible at a 3$\%$ level.}
\end{figure}
The study is based on 10 samples of MPPCs, 5 with 400 pixels and 5 with 1600 pixels. They are protected by a special plastic package. The suggested operation voltage is respectively 76~V and 78.1~V (estimated $\sim$2.5~V  and $\sim$1.4~V above the breakdown voltage), with a spread of 0.1~V. In this measurement two $3\times 3\times0.5$~cm$^{3}$ plastic organic scintillator tiles (produced by \textit{Uniplast} enterprise in Vladimir, Russia) are used. In one a 1~mm diameter green wavelength shifting fiber (Kuraray multicladding WLS fiber Y11(200)) is installed. The scintillators are wrapped in a Super-radiant VN2000 foil (3M). 

A special setup is designed in order to guarantee the reproducibility of the measurement (Fig. \ref{DR}). The scintillator is installed in a fixed housing, in the middle of a robust main structure. The MPPC is secured at one end of a plastic bar. Two grooves are carved in the main structure, one starting from the middle of the side of the tile, one from the center and they host the MPPC holder. This allows the MPPC to be coupled to the scintillator in two positions. In the direct readout setup, a window of $3\times 3$~mm$^{2}$ is open in the reflective coating, in front of the MPPC. Its holder is then completely guided in the casing and the MPPC plastic coverage  is in contact with the scintillator; this improves the coupling between scintillator and photodetector. The distance between the sensitive region of the MPPC and the surface of the scintillator is $\sim$ 2~mm. In the green wavelength shifter mediated design, the scintillator equipped with fiber is installed in the dedicated housing and the green light is directly read out by the MPPC, held by the same bar. In both designs, no specific optical coupling is used.
The response of the system to a  $\beta^{-}$ source (Ru$^{106}$) is investigated. A trigger plate, consisting of a $5\times 5\times 1$~cm$^{3}$ organic plastic scintillator, read out by a traditional photomultiplier tube, is installed behind the scintillator/MPPC system. The signal of the MPPC is amplified by the wide-band voltage amplifier Phillips Scientific 6954. The integration is performed by the QDC Lecroy 1182, in a gate of 80~ns, produced in coincidence with the trigger. 
The setup allows a reproduction of  the measurement to within a systematic error of $\pm 3\%$.             

\section{The results of the measurement} 
A characterization of the devices is initially performed. The 1600 pixels MPPC show a typical gain of ($2.7\pm 0.1$)$\cdot 10^5$. The dark rate at a threshold of 0.5 pixels\footnote{This threshold is chosen in  order to have the highest sensitivity to the dark pulses}  is ($40 \pm 1$) kHz and the cross talk ($4.3\pm 1$)$\%$. These values are measured  at 2.5~V over the breakdown voltage. The 400 pixels MPPC exhibits a gain of ($7.7\pm0.1$)$\cdot 10^5$, a dark rate at a threshold of 0.5 pixels of  ($230 \pm 3$)~kHz and cross talk of ($3.5\pm 1$)$\%$, operated at 1.4~V over the breakdown voltage. The values are found to be in agreement with the data sheet performances of the photodetector.  

\begin{figure}[h]
\centering
\subfigure[]
{\includegraphics[scale=0.4]{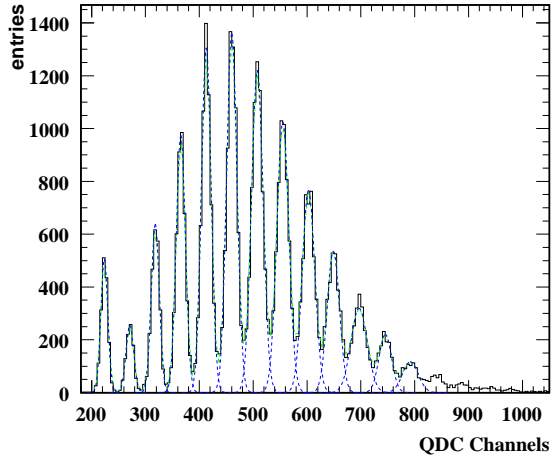}}
\subfigure[]
{\includegraphics[scale=0.4]{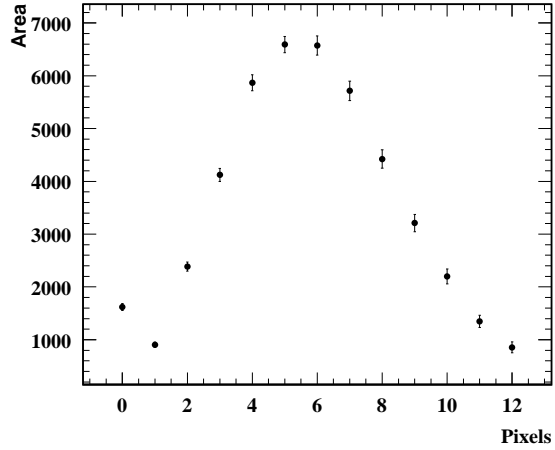}}

\caption{\label{Signal} A plastic scintillator detects a mip: it is directly read out by a 1600 pixels MPPC, operated at 2 V over the breakdown voltage. To extract the MPV of the distribution the spectrum is fit with a multi-Gaussian function (a) and the areas of the Gaussians are plotted (b). The maximum of the area plot, estimated with a Gaussian fit, is the MPV.}
\end{figure}

The aim of the experiment is to extract the most probable value (MPV) of the number of  photoelectrons produced in the MPPC when a minimum ionizing particle crosses the scintillator. The typical signal is shown in figure \ref{Signal}. Each peak corresponds to a certain number of pixels firing in the MPPC. The good separation of the peaks indicates a good uniformity of the device. 
The signal is fit with a multi-Gaussian function (Fig. \ref{Signal}.a). The extracted area of each peak is then plotted versus the number of photoelectrons (Fig. \ref{Signal}.b).  The areas follow a Landau distribution, smeared by the Poisson photo-statistics. The maximum of this distribution is the searched MPV, which is estimated with a Gaussian fit around the peak.
The results of these measurements for the green and blue light coupling for both 1600 pixels  and 400 pixels MPPC are shown in figure 3, as a function of the voltage.
\begin{figure}[h]
\centering
\subfigure[]
{\includegraphics[scale=0.4]{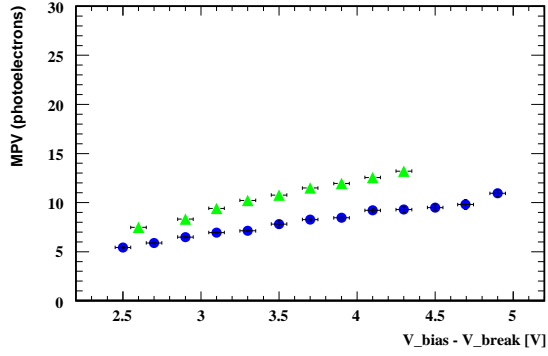}}
\subfigure[]
{\includegraphics[scale=0.4]{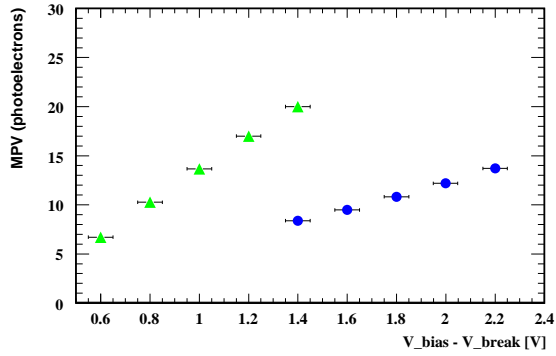}}
\caption{\label{res} Most probable value of a mip spectrum detected by a 1600 pixels (a) and a 400 pixels (b) MPPC. The blue dots correspond to the direct coupling, while the green triangles to the wavelength shifting mediated readout.}
\end{figure}

The efficiency of the 1600 pixels MPPC is found to be about half of that of the 400 pixels device, in agreement with the photo-detection efficiency declared in the producer data sheet. The MPV is in agreement within $\pm 3\%$, in both cases when the MPPC is either located centrally on the edge or near the corner. The two measurements agree within the systematic error quoted above.   

\section{Application to the calorimeter}
\begin{figure}[h]
\centering
\subfigure[]
{\includegraphics[scale=0.3]{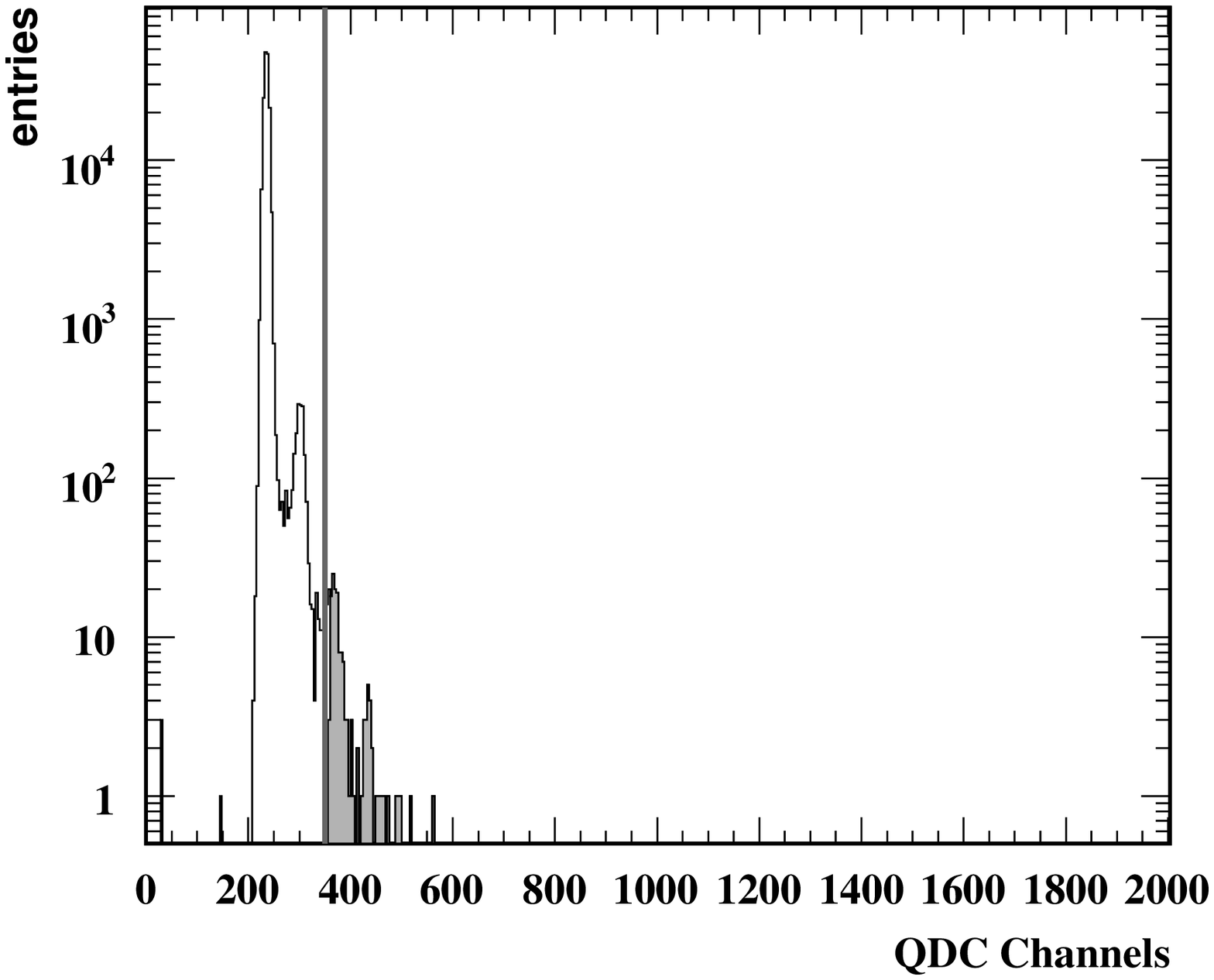}}
\subfigure[]
{\includegraphics[scale=0.3]{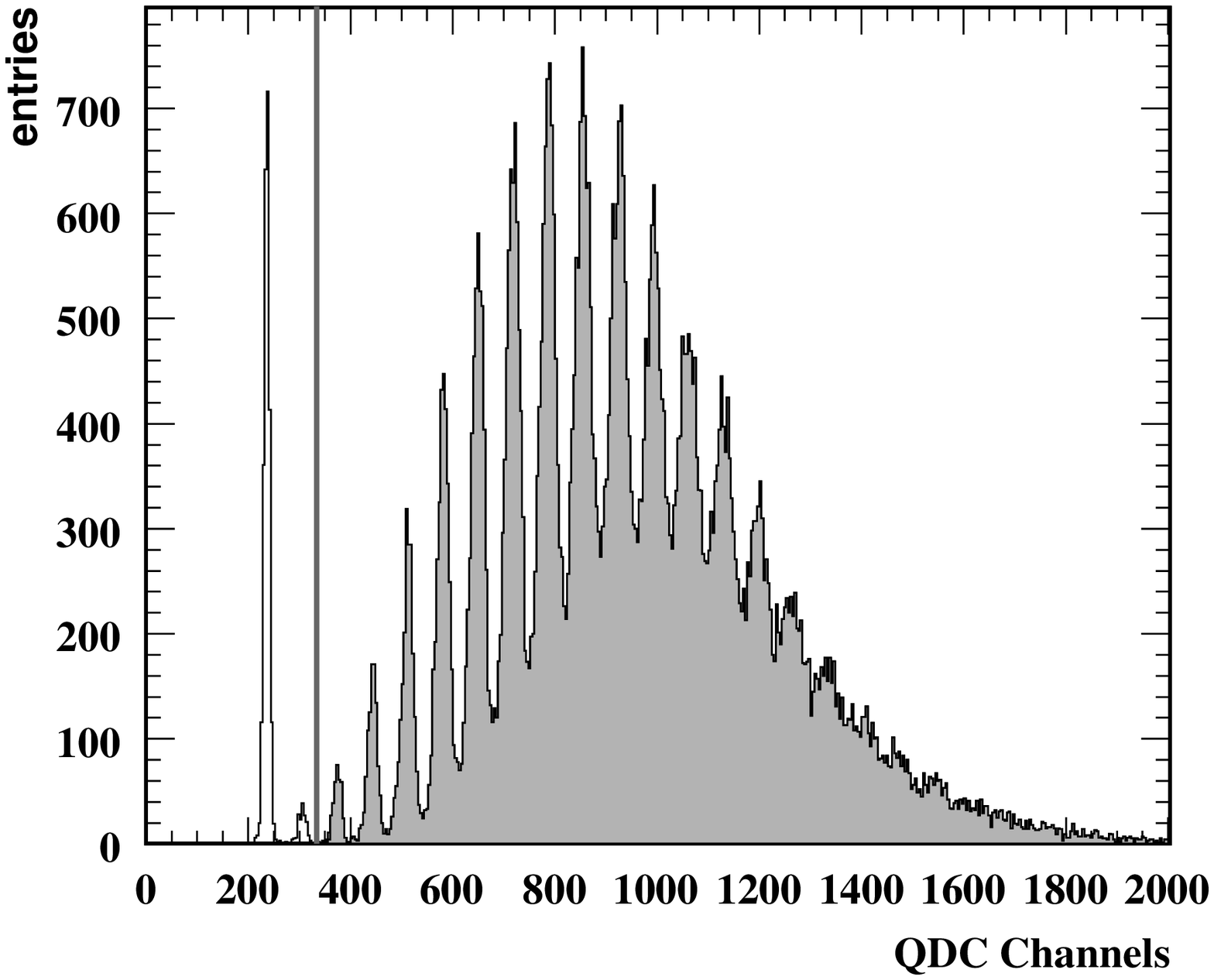}}

\caption{\label{calor} Scintillator crystal read out by a 400 pixels MPPC. A threshold is defined considerting the pedestal spectrum, so that the noise above it is 3~kHz (grey area in (a)). This threshold in amplitude defines the efficiency of the collection of the signal (b): in grey is the part of the signal above the cut, which is $98\%$ of the total, after subtracting the pedestal contribution.}
\end{figure}

In the calorimeter, mip signals are used as a reference for the calibration: the MPV of the signal sets the energy scale of each channel. An amplitude threshold\footnote{The total charge of the signal is considered as amplitude in this application.} defines the discrimination between the noise and the physics signal. The normalized integral of the mip signal above the noise threshold is the mip detection efficiency. All signals, with amplitude above threshold, constitute a hit in the calorimeter.  
The procedure is shown in figure \ref{calor}. The threshold is fixed considering the pedestal spectrum (Fig. \ref{calor}.a) such that the noise rate above threshold is 3~kHz. This requirement corresponds to an occupancy of about 8 accidental hits, as observed in the present 8000 channel prototype with $\sim 200$~ns integration time. For the ILC detector, the occupancy of $10^{-4}$ during one beam crossing interval ($\sim 300$~ns) translates into a sharper requirement of 300 Hz noise above threshold for a single photo-sensor. 

\begin{figure}[h]
\centering
\subfigure[]
{\includegraphics[scale=0.4]{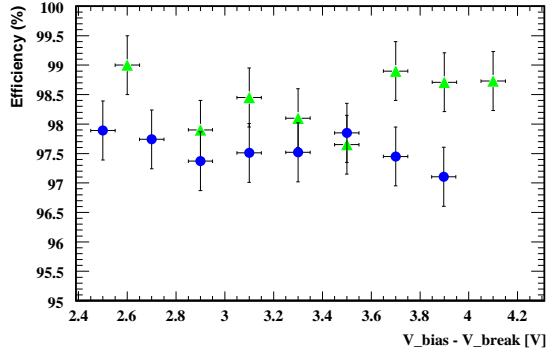}}
\subfigure[]
{\includegraphics[scale=0.4]{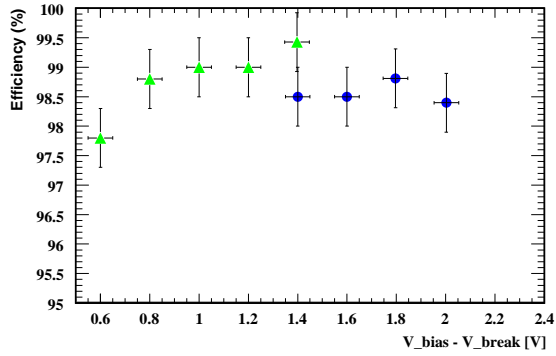}}

\caption{\label{efficiency}Signal collection efficiency for the 1600 pixels (a) and 400 pixels (b) MPPCs. The blue dots correspond to the direct coupling, while the green triangles refer to the wavelength shifting mediated readout.}
\end{figure}

The efficiency of the mip signal is then calculated using the determined threshold (\ref{calor}.b). For a given occupancy requirement, the dark noise of the sensor, the inter-pixel cross-talk and the most probable number of pixels corresponding to a mip determine the mip detection efficiency. The first defines the minimum allowed threshold, the latter determines position and Poisson width of the mip signal distribution and thus the integral above threshold. In the prototype of the hadron calorimeter, the MEPHI SiPMs provide $15\pm2$ photoelectrons at the MPV of the mip signal. The efficiency of the signal collection, as measured in the prototype,  is 95$\%$. The efficiency of the mip signal collection of MPPC, as a function of voltages, is shown in figure \ref{efficiency}. The measurements with direct coupling and with wavelength shifting fiber mediated readout are presented for both 1600 and 400 pixels MPPC.  
The MPPC have a very low dark rate and cross talk, and the requested noise cut can be put at 1.5~-~2 pixels. The 1600 pixels device, in the direct coupling option, can be used at 2.5~-~3.5~V over the breakdown, providing 6~-~7 photoelectrons at the MPV, with a collection efficiency of $\sim 97$~$\%$. The performance of this solution is then comparable with the one used in the actual prototype, and allows even an improvement in the dynamic range. If the green wavelength shifting fiber is used, the MPPC are  definitely competitive with the SiPMs in use in the present prototype: when operated at 3~-~3.5~V above the breakdown voltage, they provide a better signal efficiency ($98\%$) and a larger dynamic range.       
The 400 pixels MPPC, in both direct and wavelength shifter mediated readout, shows a very high mip signal collection efficiency, but the reduced number of pixels imposes strict bounds on the dynamic range. 

It has to be noted that in both MPPCs the dark rate drops rapidly as the threshold is raised, such that for the tighter requirements of the ILC a threshold of 2-4 pixels would be enough to keep the occupancy small. In this case it is possible to operate at a overvoltage which preserves the mip efficiency above $95\%$. If thinner scintillator, e.g. 3 mm instead of 5 mm thickness with correspondingly smaller light yield, is to be used, a better coupling or a larger sensitive area of the photodetector have to be investigated. A possible solution could be the application of the new $3\times 3$~mm$^{2}$ MPPC, but further systematical studies are needed.     

\section {Conclusion}
The direct readout of a $3\times3$~cm$^{2}$ plastic organic scintillator by a Multi Pixel Photon Counter is possible, and can be an elegant solution for the future prototype of a hadron calorimeter. The MPPCs are suitable also for the green wavelength shifter fiber readout, chosen in the actual calorimeter prototype for the ILC, providing potentially even better performances than the ones in use at the moment, in term of dynamic range and mip signal collection efficiency. The 1600 pixels devices are preferred for this particular application, as the linearity of the signal would be extended to a wider range of light intensity. The uniformity of the light collection with respect to the particle impact position on the tile  has still to be studied as well as the bounds on the dynamic range imposed by the physics observed at the linear~collider. 

A further implication of these results in the design of a Positron Emission Tomography scanner is also under study and it will be shown in detail in a future paper. 

\ack
This work is supported by DESY, by the University of Hamburg and by the Helmholtz-Nachwuchsgruppen fund VH-NG-206.
We thank V. Korbel, V. Saveliev and F. Sefkow for their useful suggestions and comments. We thank Hamamatsu, which kindly provided us the tested samples of MPPC. We thank also Peter Smirnov for his professional technical support. We finally thank all the students, from the university and the high school who contributed to this project, in particular Franziska Klingberg and Britta Riechmann from the Hamburg University, who performed an excellent laboratory practice at DESY.

\end{document}